\pdfoutput=1

\documentclass[english]{article}
\usepackage[T1]{fontenc}
\usepackage[latin9]{inputenc}
\usepackage{babel}
\usepackage{amsmath}
\usepackage{amssymb}
\usepackage{stmaryrd}
\usepackage{xargs}[2008/03/08]
\usepackage[unicode=true,pdfusetitle,
 bookmarks=true,bookmarksnumbered=false,bookmarksopen=false,
 breaklinks=false,pdfborder={0 0 0},backref=false,colorlinks=false]
 {hyperref}

\makeatletter

\newcommand{\mathcircumflex}[0]{\mbox{\^{}}}

\usepackage{bussproofs}
\usepackage{centernot}
\usepackage{datetime}
\usepackage{filecontents}
\usepackage{fullpage}
\usepackage{lmodern}
\usepackage{microtype}
\usepackage{tikz}
\usepackage{pgfplots}

\usetikzlibrary{automata}
\usetikzlibrary{backgrounds}
\usetikzlibrary{fit}
\usetikzlibrary{positioning}

\makeatother

\usepackage{listings}

\begin{document}

\title{Language to Specify Syntax-Guided Synthesis Problems}

\author{Mukund Raghothaman \and Abhishek Udupa}

\maketitle

\global\long\def\autobox#1{#1}

\global\long\def\opname#1{\autobox{\operatorname{#1}}}

\global\long\def\dontcare{\_}

\global\long\def\bbracket#1{\left\llbracket #1\right\rrbracket }

\global\long\def\cnot#1{\centernot#1}

\global\long\def\roset#1{\left\{  #1\right\}  }

\global\long\def\ruset#1#2{\roset{#1\;\middle\vert\;#2}}

\global\long\def\union#1#2{#1\cup#2}

\global\long\def\bigunion#1#2{\bigcup_{#1}#2}

\global\long\def\intersection#1#2{#1\cap#2}

\global\long\def\bigintersection#1#2{\bigcap_{#1}#2}

\global\long\def\bigland#1#2{\bigwedge_{#1}#2}

\global\long\def\powerset#1{2^{#1}}

\global\long\def\cart#1#2{#1\times#2}

\global\long\def\tuple#1{\left(#1\right)}

\global\long\def\quoset#1#2{\left.#1\middle/#2\right.}

\global\long\def\equivclass#1{\left[#1\right]}

\global\long\def\refltransclosure#1{#1^{*}}

\newcommandx\funcapphidden[3][usedefault, addprefix=\global, 1=]{#2#1#3}

\global\long\def\funcapplambda#1#2{\funcapphidden[\,]{#1}{#2}}

\global\long\def\funcapptrad#1#2{\funcapphidden{#1}{\tuple{#2}}}

\global\long\def\funccomp#1#2{#1\circ#2}

\global\long\def\arrow#1#2{#1\to#2}

\global\long\def\func#1#2#3{#1:\arrow{#2}{#3}}

\global\long\def\N{\mathbb{N}}

\global\long\def\Z{\mathbb{Z}}

\global\long\def\Q{\mathbb{Q}}

\global\long\def\R{\mathbb{R}}

\global\long\def\D{\mathbb{D}}

\global\long\def\Zmod#1{\quoset{\Z}{#1\Z}}

\global\long\def\vector#1{\mathbf{#1}}

\global\long\def\dotprod#1#2{#1\cdot#2}

\global\long\def\bool{{\tt bool}}

\global\long\def\true{{\tt true}}

\global\long\def\false{{\tt false}}

\global\long\def\strempty{\epsilon}

\global\long\def\kstar#1{#1^{*}}

\global\long\def\kplus#1{#1^{+}}

\global\long\def\paren#1{{\tt (}#1{\tt )}}

\global\long\def\sygus{\left\langle SyGuS\right\rangle }

\global\long\def\cmd{\left\langle Cmd\right\rangle }

\global\long\def\setlogiccmd{\left\langle SetLogicCmd\right\rangle }

\global\long\def\sortdefcmd{\left\langle SortDefCmd\right\rangle }

\global\long\def\fundefcmd{\left\langle FunDefCmd\right\rangle }

\global\long\def\vardeclcmd{\left\langle VarDeclCmd\right\rangle }

\global\long\def\fundeclcmd{\left\langle FunDeclCmd\right\rangle }

\global\long\def\synthfuncmd{\left\langle SynthFunCmd\right\rangle }

\global\long\def\constraintcmd{\left\langle ConstraintCmd\right\rangle }

\global\long\def\checksynthcmd{\left\langle CheckSynthCmd\right\rangle }

\global\long\def\setoptscmd{\left\langle SetOptsCmd\right\rangle }

\global\long\def\symbol{\left\langle Symbol\right\rangle }

\global\long\def\quotedliteral{\left\langle QuotedLiteral\right\rangle }

\global\long\def\sortexpr{\left\langle SortExpr\right\rangle }

\global\long\def\intconst{\left\langle IntConst\right\rangle }

\global\long\def\realconst{\left\langle RealConst\right\rangle }

\global\long\def\boolconst{\left\langle BoolConst\right\rangle }

\global\long\def\bvconst{\left\langle BVConst\right\rangle }

\global\long\def\enumconst{\left\langle EnumConst\right\rangle }

\global\long\def\literal{\left\langle Literal\right\rangle }

\global\long\def\term{\left\langle Term\right\rangle }

\global\long\def\gterm{\left\langle GTerm\right\rangle }

\global\long\def\letterm{\left\langle LetTerm\right\rangle }

\global\long\def\letgterm{\left\langle LetGTerm\right\rangle }

\global\long\def\ntdef{\left\langle NTDef\right\rangle }

\global\long\def\setlogickwd{{\tt set\mbox{-}logic}}

\global\long\def\sortdefkwd{{\tt define\mbox{-}sort}}

\global\long\def\fundefkwd{{\tt define\mbox{-}fun}}

\global\long\def\vardeclkwd{{\tt declare\mbox{-}var}}

\global\long\def\fundeclkwd{{\tt declare\mbox{-}fun}}

\global\long\def\synthfunkwd{{\tt synth\mbox{-}fun}}

\global\long\def\constraintkwd{{\tt constraint}}

\global\long\def\checksynthkwd{{\tt check\mbox{-}synth}}

\global\long\def\setoptskwd{{\tt set\mbox{-}options}}

\global\long\def\bitveckwd{{\tt BitVec}}

\global\long\def\arraykwd{{\tt Array}}

\global\long\def\intkwd{{\tt Int}}

\global\long\def\boolkwd{{\tt Bool}}

\global\long\def\enumkwd{{\tt Enum}}

\global\long\def\realkwd{{\tt Real}}

\global\long\def\constantkwd{{\tt Constant}}

\global\long\def\varkwd{{\tt Variable}}

\global\long\def\inputvarkwd{{\tt InputVariable}}

\global\long\def\localvarkwd{{\tt LocalVariable}}

\global\long\def\letkwd{{\tt let}}

\global\long\def\truekwd{{\tt true}}

\global\long\def\falsekwd{{\tt false}}

\section{Introduction \label{sec:Intro}}

We present a language to specify syntax guided synthesis (SyGuS) problems.
Syntax guidance is a prominent theme in contemporary program synthesis
approaches, and SyGuS was first described in \cite{FMCAD13}. An instance
of a SyGuS problem has four parts:
\begin{enumerate}
\item A base vocabulary and theory, specifying the basic types, primitive
operations over the types, and their properties,
\item a finite set of typed ``synthesis'' functions $f_{1}$, $f_{2}$,
\ldots{}, whose bodies are to be synthesized,
\item syntactic constraints: for each synthesis function $f_{i}$, a grammar
$G_{i}$ describing the syntactic structure of the potential solutions,
and
\item semantic constraints: a formula $\varphi$, with some universally
quantified variables $v_{1}$, $v_{2}$, \ldots{}, which constrains
the values of the synthesis functions.
\end{enumerate}
The problem is to find expression bodies for each synthesis function
$f_{i}$ from the grammar $G_{i}$ so that the constraint is universally
satisfied:

\begin{alignat*}{1}
 & \forall v_{1},v_{2},\ldots,\funcapptrad{\varphi}{f_{1},f_{2},\ldots,v_{1},v_{2},\ldots}.
\end{alignat*}
The constraint formula $\varphi$ is quantifier-free, and the logical
symbols and their interpretation in $\varphi$ and the grammar are
restricted to a background theory.

For example, over the theory of linear integer arithmetic, the functions
computing the maximum $\autobox{max}_{2}$ and minimum $\autobox{min}_{2}$
of a pair of integers may be specified as 
\begin{alignat*}{1}
 & \begin{array}{ll}
\forall x,y:\Z, & \funcapptrad{\autobox{max}_{2}}{x,y}\geq x\land\funcapptrad{\autobox{max}_{2}}{x,y}\geq y\\
 & \land\left(\funcapptrad{\autobox{max}_{2}}{x,y}=x\lor\funcapptrad{\autobox{max}_{2}}{x,y}=y\right)\\
 & \land\left(\funcapptrad{\autobox{max}_{2}}{x,y}+\funcapptrad{\autobox{min}_{2}}{x,y}=x+y\right).
\end{array}
\end{alignat*}

We are interested in piecewise linear functions, so the grammar $G$
for both functions would be 
\begin{alignat*}{1}
 & \begin{array}{rcl}
{\tt Expr} & ::= & \begin{array}{lllllll}
0 & | & 1 & | & x & | & y\end{array}\\
 & | & {\tt Expr}+{\tt Expr}\\
 & | & {\tt Expr}-{\tt Expr}\\
 & | & \paren{{\tt ite}\mbox{ }{\tt BoolExpr}\mbox{ }{\tt Expr}\mbox{ }{\tt Expr}}\\
{\tt BoolExpr} & ::= & {\tt BoolExpr}\land{\tt BoolExpr}\\
 & | & \lnot{\tt BoolExpr}\\
 & | & {\tt Expr}\leq{\tt Expr}
\end{array}
\end{alignat*}

\section{Example SyGuS Specification \label{sec:Example}}

Before formally describing the language, we present a concrete example
of a SyGuS specification.

\begin{figure}
\begin{lstlisting}[basicstyle={\ttfamily}]
(set-logic LIA)

(synth-fun max2 ((x Int) (y Int)) Int
   ((Start Int (0 1 x y
                (+ Start Start)
                (- Start Start)
                (ite StartBool Start Start)))

    (StartBool Bool ((and StartBool StartBool)
                     (not StartBool)
                     (<=  Start Start)))))

(synth-fun min2 ((x Int) (y Int)) Int
   ((Start Int ((Constant Int) (Variable Int)
                (+ Start Start)
                (- Start Start)
                (ite StartBool Start Start)))

    (StartBool Bool ((and StartBool StartBool)
                     (not StartBool)
                     (<=  Start Start)))))

(declare-var x Int)
(declare-var y Int)

(constraint (>= (max2 x y) x))
(constraint (>= (max2 x y) y))

(constraint (or (= x (max2 x y))
            (or (= y (max2 x y)))))

(constraint (= (+ (max2 x y) (min2 x y))
               (+ x y)))

(check-synth)
\end{lstlisting}

\caption{SyGuS specification for functions computing the maximum and minimum
of two integers. \label{fig:Example:Max2}}
\end{figure}

We continue the example of $\autobox{max}_{2}$ and $\autobox{min}_{2}$
from the previous section, and present the corresponding SyGuS code
in figure \ref{fig:Example:Max2}. The first command $\paren{\setlogickwd\mbox{ }{\tt LIA}}$
informs the synthesizer to load symbols corresponding to linear integer
arithmetic. Next, we describe the functions to be synthesized: the
command $\paren{\synthfunkwd\mbox{ }{\tt max2}\mbox{ }\ldots}$ command
first specifies that ${\tt max2}$ is a function of two integer arguments
${\tt x}$ and ${\tt y}$, and returns an integer value. The rest
of the command describes the grammar for ${\tt max2}$. ${\tt Start}$
and ${\tt StartBool}$ are integer-valued and boolean-valued non-terminal
symbols respectively. ${\tt Start}$ is the special starting non-terminal
of the grammar. The description of ${\tt min2}$ is identical to that
of ${\tt max2}$, except for the function name, and some useful shorthands
$\paren{\constantkwd\mbox{ }\intkwd}$ and $\paren{\varkwd\mbox{ }\intkwd}$
which respectively expand to any integer constant and integer-valued
variable currently in scope. Finally, the code lists the constraints
that these functions satisfy. Pick a pair of integers ${\tt x}$ and
${\tt y}$. The first constraint requires that $\funcapptrad{{\tt max2}}{{\tt x},{\tt y}}\geq{\tt x}$.
The final synthesis contraint $\varphi$ is the conjunction of the
constraints imposed by the individual constraint commands.

\section{Specification Language \label{sec:Spec}}

The SyGuS specification language is closely modeled on SMT-Lib2. A
SyGuS input file is a sequence of commands; in subsections \ref{sub:Spec:SetLogic}-\ref{sub:Spec:SetOpts},
we describe the syntax of each command. In the following description,
italicized text within angle-brackets represents EBNF non-terminals,
and text in typewriter font represents terminal symbols.

\begin{alignat*}{1}
 & \begin{array}{rcl}
\sygus & ::= & \setlogiccmd\kplus{\cmd}\\
 & | & \kplus{\cmd}\\
\cmd & ::= & \sortdefcmd\\
 & | & \vardeclcmd\\
 & | & \fundeclcmd\\
 & | & \fundefcmd\\
 & | & \synthfuncmd\\
 & | & \constraintcmd\\
 & | & \checksynthcmd\\
 & | & \setoptscmd
\end{array}
\end{alignat*}

\subsection{Language trivia \label{sub:Spec:Trivia}}

\subsubsection{Reserved words \label{sub:Spec:Trivia:Keywords}}

The following keywords are reserved, and may not be used as identifiers
in any context: $\setlogickwd$, $\sortdefkwd$, $\vardeclkwd$, $\fundeclkwd$,
$\fundefkwd$, $\synthfunkwd$, $\constraintkwd$, $\checksynthkwd$,
$\setoptskwd$, $\bitveckwd$, $\arraykwd$, $\intkwd$, $\boolkwd$,
$\enumkwd$, $\realkwd$, $\constantkwd$, $\varkwd$, $\inputvarkwd$,
$\localvarkwd$, $\letkwd$, $\truekwd$, $\falsekwd$.

\subsubsection{Comments \label{sub:Spec:Trivia:Comments}}

Comments in SyGuS specifications are indicated by a semicolon ${\tt ;}$.
On encountering a ${\tt ;}$, the rest of the line is ignored.

\subsubsection{Identifiers \label{sub:Spec:Trivia:Identifiers}}

Identifiers are denoted with the non-terminal $\symbol$. An identifier
is any non-empty sequence of upper- and lower-case alphabets, digits,
and certain special characters, with the restriction that it may not
begin with a digit.

\begin{alignat*}{1}
 & \begin{array}{rcl}
\left\langle SpecialChar\right\rangle  & = & \roset{{\tt \_},{\tt +},{\tt -},{\tt *},{\tt \&},{\tt |},{\tt !},{\tt \sim},{\tt <},{\tt >},{\tt =},{\tt /},{\tt \%},{\tt ?},{\tt .},{\tt \$},{\tt \mathcircumflex}}\\
\symbol & ::= & \left(\begin{array}{lllll}
\left[{\tt a}-{\tt z}\right] & | & \left[{\tt A}-{\tt Z}\right] & | & \left\langle SpecialChar\right\rangle \end{array}\right)\\
 &  & \kstar{\left(\begin{array}{ccccccc}
\left[{\tt a}-{\tt z}\right] & | & \left[{\tt A}-{\tt Z}\right] & | & \left[{\tt 0}-{\tt 9}\right] & | & \left\langle SpecialChar\right\rangle \end{array}\right)}
\end{array}
\end{alignat*}

A quoted literal, $\left\langle QuotedLiteral\right\rangle $ is a
non-empty sequence of alphabets, digits and the period (${\tt .}$)
enclosed within double-quotes.

\begin{alignat*}{1}
 & \begin{array}{rcl}
\left\langle QuotedLiteral\right\rangle  & ::= & \texttt{"}\kplus{\left(\begin{array}{ccccccc}
\left[{\tt a}-{\tt z}\right] & | & \left[{\tt A}-{\tt Z}\right] & | & \left[{\tt 0}-{\tt 9}\right] & | & {\tt .}\end{array}\right)}\texttt{"}\end{array}
\end{alignat*}

\subsubsection{Literals \label{sub:Spec:Trivia:Literals}}

\begin{alignat*}{1}
 & \begin{array}{rcl}
\literal & ::= & \begin{array}{ccccc}
\intconst & | & \realconst & | & \boolconst\end{array}\\
 & | & \begin{array}{ccc}
\bvconst & | & \enumconst\end{array}\\
\intconst & ::= & \begin{array}{ccc}
\kplus{\left[{\tt 0}-{\tt 9}\right]} & | & {\tt -}\kplus{\left[{\tt 0}-{\tt 9}\right]}\end{array}\\
\realconst & ::= & \begin{array}{ccc}
\kplus{\left[{\tt 0}-{\tt 9}\right]}{\tt .}\kplus{\left[{\tt 0}-{\tt 9}\right]} & | & {\tt -}\kplus{\left[{\tt 0}-{\tt 9}\right]}{\tt .}\kplus{\left[{\tt 0}-{\tt 9}\right]}\end{array}\\
\boolconst & ::= & \begin{array}{ccc}
\truekwd & | & \falsekwd\end{array}\\
\bvconst & ::= & \begin{array}{ccc}
{\tt \#b}\kplus{\left[{\tt 0}-{\tt 1}\right]} & | & {\tt \#x}\kplus{\left(\begin{array}{ccccc}
\left[{\tt 0-9}\right] & | & \left[{\tt a}-{\tt f}\right] & | & \left[{\tt A}-{\tt F}\right]\end{array}\right)}\end{array}\\
\enumconst & ::= & \symbol{\tt ::}\symbol
\end{array}
\end{alignat*}

Integer constants are written as usual, in decimal, with an optional
minus at the beginning to denote a negative number. Real numbers are
written using their decimal expansion: at least one decimal digit
before and after a mandatory period, and an optional minus sign at
the beginning. $\truekwd$ and $\falsekwd$ are the predefined boolean
constants. Bit-vector constants may be written using either their
traditional binary or hexadecimal representations. Enumerated constants
are written in two parts: the first identifier names the sort the
constant belongs to, and the second identifier names the constructor.
The definition of enumerated sorts is described in subsection \ref{sub:Spec:SortDef}.

\subsection{Declaring the problem logic \label{sub:Spec:SetLogic} \hfill{}
$\protect\setlogiccmd$}

On encountering the optional $\setlogiccmd$, the synthesizer loads
appropriate pre-defined function symbols and constants. Current theories
include
\begin{enumerate}
\item ${\tt LIA}$: Linear integer arithmetic, for functions such as ${\tt +}$
and ${\tt -}$,
\item ${\tt BV}$: Theory of bit-vectors, for functions such as ${\tt bvadd}$
and ${\tt bvlshr}$,
\item ${\tt Reals}$: Theory of real numbers, and
\item ${\tt Arrays}$: Theory of arrays.
\end{enumerate}
\begin{alignat*}{1}
 & \begin{array}{rcl}
\setlogiccmd & ::= & \paren{\setlogickwd\mbox{ }\symbol}\end{array}
\end{alignat*}

\subsection{Defining new sorts \label{sub:Spec:SortDef} \hfill{} $\protect\sortdefcmd$,
$\protect\sortexpr$}

SyGuS expects that the sorts of functions, variables, and grammar
symbols be explicitly specified. The syntactic construct $\sortexpr$
is used for this, and the sort definition command $\sortdefcmd$ permits
defining useful shorthands.

\begin{alignat*}{1}
 & \begin{array}{rcl}
\sortexpr & ::= & \begin{array}{lllll}
\intkwd & | & \boolkwd & | & \realkwd\end{array}\\
 & | & \paren{\bitveckwd\mbox{ }\left\langle PositiveInteger\right\rangle }\\
 & | & \paren{\enumkwd\mbox{ }\paren{\kplus{\symbol}}}\\
 & | & \paren{\arraykwd\mbox{ }\sortexpr\mbox{ }\sortexpr}\\
 & | & \symbol\\
\sortdefcmd & ::= & \paren{\sortdefkwd\mbox{ }\symbol\mbox{ }\sortexpr}
\end{array}
\end{alignat*}

The sorts $\intkwd$, $\boolkwd$, and $\realkwd$ refer to integers,
booleans and real numbers respectively. For each positive integer
$n$, $\bitveckwd\mbox{ }n$ refers to the sort of bit-vectors $n$
bits long. Given a set of constructor symbols $S_{1}$, $S_{2}$,
\ldots{}, the sort $\paren{\enumkwd\mbox{ }\paren{S_{1}\mbox{ }S_{2}\mbox{ }\ldots}}$
refers to the enumerated type having those elements. Since the only
way to represent an enumerated constant (subsection \ref{sub:Spec:Trivia:Literals})
is by also specifying the sort-name, the constructors $S_{1}$, $S_{2}$
etc. may have the same names as previously defined variables, functions,
or sorts. The sort $\paren{\arraykwd\mbox{ }S_{1}\mbox{ }S_{2}}$
represents arrays that map elements of sort $S_{1}$ to elements of
sort $S_{2}$.

Once a sort $S$ has been defined using the command $\paren{\sortdefkwd\mbox{ }S\mbox{ }\sortexpr}$,
it may subsequently be referred to simply as $S$ rather than the
full expression $\sortexpr$. The identifier $\symbol$ used to name
a sort should not have been previously used as a sort name. Every
$\sortexpr$ in a SyGuS specification must be well-formed. We say
that a $\sortexpr$ is well-formed if
\begin{enumerate}
\item it is an instance of $\intkwd$, $\boolkwd$, $\realkwd$, $\bitveckwd$
or $\enumkwd$, or
\item it is an instance of $\arraykwd$ and both domain and range of the
array sort are well-formed, or
\item it is a $\symbol$, and $\symbol$ has been previously defined using
a $\sortdefcmd$.
\end{enumerate}

\subsection{Universally quantified variables \label{sub:Spec:VarDecl} \hfill{}
$\protect\vardeclcmd$}

Universally quantified variables may be declared with $\vardeclcmd$.
\begin{alignat*}{1}
 & \begin{array}{rcl}
\vardeclcmd & ::= & \paren{\vardeclkwd\mbox{ }\symbol\mbox{ }\sortexpr}\end{array}
\end{alignat*}

The variable name $\symbol$ must not clash with the following:
\begin{enumerate}
\item any previously declared universally quantified variable ($\vardeclcmd$), 
\item any previously declared $0$-arity uninterpreted function ($\fundeclcmd$), 
\item any previously defined $0$-arity function macro ($\fundefcmd$),
and
\item any previously declared $0$-arity synthesis function ($\synthfuncmd$).
\end{enumerate}

\subsection{Uninterpreted functions \label{sub:Spec:Uninterp} \hfill{} $\protect\fundeclcmd$}

Uninterpreted functions are declared using $\fundeclcmd$.

\subsubsection{Syntax \label{sub:Spec:Uninterp:Syntax}}

\begin{alignat*}{1}
 & \begin{array}{rcl}
\fundeclcmd & ::= & \paren{\fundeclkwd\mbox{ }\symbol\mbox{ }\paren{\kstar{\sortexpr}}\mbox{ }\sortexpr}\end{array}
\end{alignat*}

The $\symbol$ names the uninterpreted function being declared, the
first list of $\sortexpr$ identifies the number and sorts of the
input arguments, and the final $\sortexpr$ identifies the sort of
the function return value. The function name $\symbol$ must not clash
with the following:
\begin{enumerate}
\item if the funtion is of $0$-arity, then $\symbol$ should not clash
with any previously declared universally quantified variable ($\vardeclcmd$),
\item any previously declared uninterpreted function ($\fundeclcmd$) with
the same input argument type signature,
\item any previously defined function macro ($\fundefcmd$) with the same
input argument type signature, and
\item any previously declared synthesis function ($\synthfuncmd$) with
the same input argument type signature.
\end{enumerate}

\subsubsection{Semantics \label{sub:Spec:Uninterp:Semantics}}

When uninterpreted functions are used in a SyGuS problem, the synthesized
functions must satisfy the specification for all models of the uninterpreted
functions. Uninterpreted functions may only be used in constraints
(section \ref{sub:Spec:Constraint}), and not in function macros or
grammars (sections \ref{sub:Spec:FunDef} and \ref{sub:Spec:SynthFun}).

For example, consider the specification in figure \ref{fig:Spec:Uninterp:Semantics:Ex}.
Informally, this requires that for all functions $\func{{\tt uf}}{\Z}{\Z}$
and integers $x\in\Z$, $\funcapptrad{{\tt f}}{\funcapptrad{{\tt uf}}x,\funcapptrad{{\tt uf}}x}$
must hold. Therefore, the function in figure \ref{fig:Spec:Uninterp:Semantics:ExAns}
satisfies the specification, but the function in figure \ref{fig:Spec:Uninterp:Semantics:ExNonAns}
does not, even though it works for a specific instance of ${\tt uf}$,
viz. $\forall x\in\Z$, $\funcapptrad{{\tt uf}}x=5$.

\begin{figure}
\begin{lstlisting}[basicstyle={\ttfamily}]
(set-logic LIA)

(declare-fun uf (Int) Int)

(synth-fun f ((x Int) (y Int)) Bool
   ((Start Bool (true false
                 (<= IntExpr IntExpr)
                 (= IntExpr IntExpr)
                 (and Start Start)
                 (or Start Start)
                 (not Start)))
    (IntExpr Int (0 1 x y
                  (+ IntExpr IntExpr)
                  (- IntExpr IntExpr)))))

(declare-var x Int)

(constraint (f (uf x) (uf x)))

(check-synth)
\end{lstlisting}

\caption{\label{fig:Spec:Uninterp:Semantics:Ex} Example SyGuS specification
using uninterpreted functions.}
\end{figure}

\begin{figure}
\begin{lstlisting}[basicstyle={\ttfamily}]
(define-fun f ((x Int) (y Int)) Bool
   (= x y))
\end{lstlisting}

\caption{\label{fig:Spec:Uninterp:Semantics:ExAns} Sample valid answer for
SyGuS specification of figure \ref{fig:Spec:Uninterp:Semantics:Ex}.}
\end{figure}

\begin{figure}
\begin{lstlisting}[basicstyle={\ttfamily}]
(define-fun f ((x Int) (y Int)) Bool
   (= x 5))
\end{lstlisting}

\caption{\label{fig:Spec:Uninterp:Semantics:ExNonAns} Example incorrect solution
to the specification of figure \ref{fig:Spec:Uninterp:Semantics:Ex}.
Note that even though this works for some instances of ${\tt uf}$,
it is incorrect because it does not work for all.}
\end{figure}

\subsection{Terms and grammars \label{sub:Spec:TermGTerm} \hfill{} $\protect\term$,
$\protect\gterm$}

\begin{alignat*}{1}
 & \begin{array}{rcl}
\term & ::= & \paren{\symbol\mbox{ }\kstar{\term}}\\
 & | & \literal\\
 & | & \symbol\\
 & | & \letterm\\
\letterm & ::= & \paren{\letkwd\mbox{ }\paren{\kplus{\paren{\symbol\mbox{ }\sortexpr\mbox{ }\term}}}\mbox{ }\term}
\end{array}
\end{alignat*}

\begin{alignat*}{1}
 & \begin{array}{rcl}
\gterm & ::= & \paren{\symbol\mbox{ }\kstar{\gterm}}\\
 & | & \literal\\
 & | & \symbol\\
 & | & \letgterm\\
 & | & \paren{\constantkwd\mbox{ }\sortexpr}\\
 & | & \paren{\varkwd\mbox{ }\sortexpr}\\
 & | & \paren{\inputvarkwd\mbox{ }\sortexpr}\\
 & | & \paren{\localvarkwd\mbox{ }\sortexpr}\\
\letgterm & ::= & \paren{\letkwd\mbox{ }\paren{\kplus{\paren{\symbol\mbox{ }\sortexpr\mbox{ }\gterm}}}\mbox{ }\gterm}
\end{array}
\end{alignat*}

To describe function macros, grammars and constraints in SyGuS, one
uses the $\term$ and $\gterm$ constructs. The difference between
the two is the set of predefined macros (such as $\paren{\constantkwd\mbox{ }\ldots}$,
etc.) that a $\gterm$ may expand to. To allow synthesizers to perform
common subexpression elimination to speed up their computation or
reduce the size of their answers, $\letkwd$-expressions are allowed.

In grammars, a grammar expansion ${\tt (}\constantkwd\mbox{ }\sortexpr{\tt )}$
expands to any literal of type $\sortexpr$. ${\tt (}\varkwd\mbox{ }\sortexpr{\tt )}$
expands to any variable currently in score of appropriate type, ${\tt (}\inputvarkwd\mbox{ }\sortexpr{\tt )}$
and ${\tt (}\localvarkwd\mbox{ }\sortexpr{\tt )}$ expand to any formal
argument of the synthesis function, and any variable bound locally
within a $\letkwd$-expression respectively.

The interpretation of the various syntactic constructs is as usual.
In a $\letkwd$-construct, the first set of bindings ${\tt (}{\tt (}\symbol\mbox{ }\term{\tt )}^{+}{\tt )}$
(resp. $\gterm$) refers to the parallel assignment of each $\term$
(resp. $\gterm$) to the corresponding $\symbol$, as is the case
in SMT-Lib2. If the $\symbol$ bound by a $\letkwd$-expression is
already bound, then its value is shadowed while evaluating the nested
$\term$.

$\term$ and $\gterm$ constructs are type-checked in the intuitive
manner. The important restriction is that $\letkwd$-bound variables
can shadow previously declared variables only if they are of the same
sort.

\subsection{Defining macros \label{sub:Spec:FunDef} \hfill{} $\protect\fundefcmd$}

\begin{alignat*}{1}
 & \begin{array}{rcl}
\fundefcmd & ::= & \paren{\fundefkwd\mbox{ }\symbol\mbox{ }\paren{\kstar{\paren{\symbol\mbox{ }\sortexpr}}}\mbox{ }\sortexpr\mbox{ }\term}\end{array}
\end{alignat*}

$\fundefcmd$ command defines a function macro.
\begin{enumerate}
\item The function name $\symbol$ may not clash with the following:

\begin{enumerate}
\item if the funtion is of $0$-arity, then $\symbol$ should not clash
with any previously declared universally quantified variable ($\vardeclcmd$),
\item any previously declared uninterpreted function ($\fundeclcmd$) with
the same input argument type signature,
\item any previously defined function macro ($\fundefcmd$) with the same
input argument type signature, and
\item any previously declared synthesis function ($\synthfuncmd$) with
the same input argument type signature.
\end{enumerate}
\item All arguments must have distinct names.
\item No nested $\letkwd$-bound variable in $\term$ may shadow an input
argument to the function.
\item $\term$ is interpreted in the scope containing all previously defined
function macros and formal arguments.
\item The sort of $\term$ must match the return sort mentioned in $\sortexpr$.
\end{enumerate}

\subsection{Defining synthesis functions \label{sub:Spec:SynthFun} \hfill{}
$\protect\synthfuncmd$}

\begin{alignat*}{1}
 & \begin{array}{rcl}
\synthfuncmd & ::= & \paren{\synthfunkwd\mbox{ }\symbol\mbox{ }\paren{\kstar{\paren{\symbol\mbox{ }\sortexpr}}}\mbox{ }\sortexpr\mbox{ }\paren{\kplus{\ntdef}}}\\
\ntdef & ::= & \paren{\symbol\mbox{ }\sortexpr\mbox{ }\paren{\kplus{\gterm}}}
\end{array}
\end{alignat*}

A $\synthfuncmd$ describes the sort and syntax of a function to be
synthesized. The $\synthfuncmd$ specifies the function name, input
parameters, output sort, and grammar production rules respectively.
The production rules corresponding to each non-terminal are described
by an $\ntdef$, which specifies, in order, the non-terminal name,
the sort of the resulting productions, and a non-empty sequence of
production rules. Each $\gterm$ corresponds to a production rule.
\begin{enumerate}
\item The function name $\symbol$ may not clash with the following:

\begin{enumerate}
\item if the funtion is of $0$-arity, then $\symbol$ should not clash
with any previously declared universally quantified variable ($\vardeclcmd$),
\item any previously declared uninterpreted function ($\fundeclcmd$) with
the same input argument type signature,
\item any previously defined function macro ($\fundefcmd$) with the same
input argument type signature, and
\item any previously declared synthesis function ($\synthfuncmd$) with
the same input argument type signature.
\end{enumerate}
\item All arguments must have distinct names.
\item No nested $\letkwd$-bound variable in any $\gterm$ may shadow an
input argument to the function.
\item All non-terminals must have unique names. For each non-terminal, its
name should not clash with any of the following:

\begin{enumerate}
\item any previously defined $0$-arity function macro ($\fundefcmd$),
\item any formal argument to the function, and
\item any $\letkwd$-bound variable in any production rule.
\end{enumerate}
\item All $\letkwd$-bound variables in all $\gterm$s with the same name
have the same type.
\item Each production rule is interpreted in the scope with the following
in scope:

\begin{enumerate}
\item all previously defined function macros,
\item all formal arguments to the function, and
\item all $\letkwd$-bound variables in all production rules. For an example
of why this is the case, consider that the expansion ${\tt Start}\to{\tt z}$
is well-formed in the grammar of figure \ref{fig:Spec:SynthFun:Example:Let}.
\end{enumerate}
\item The sort of each production rule $\gterm$ must match the sort at
the non-terminal declaration.
\item There must be a non-terminal named ${\tt Start}$. The sort of this
non-terminal must match the ouput sort of the $\synthfuncmd$ being
declared.
\end{enumerate}
\begin{figure}
\begin{lstlisting}[basicstyle={\ttfamily}]
(synth-fun f ((x Int) (y Int)) Int
   ((Start Int (x y z
                (+ Start Start)
                (let ((z Int Start)) Start)))))
\end{lstlisting}

\caption{Example of a well-formed $\protect\synthfuncmd$ involving $\protect\letkwd$-expressions.
\label{fig:Spec:SynthFun:Example:Let}}
\end{figure}

\subsection{Describing synthesis constraints \label{sub:Spec:Constraint} \hfill{}
$\protect\constraintcmd$}

\begin{alignat*}{1}
 & \begin{array}{rcl}
\constraintcmd & ::= & \paren{\constraintkwd\mbox{ }\term}\end{array}
\end{alignat*}

A $\constraintcmd$ adds the constraint that when the synthesized
functions are substituted into $\term$, for all values of the universally
quantified variables, and all models of uniterpreted functions, $\term$
evaluates to true. $\term$ must have boolean sort in the context
with the following in scope:
\begin{enumerate}
\item all previously declared universally quantified variables,
\item all previously declared uninterpreted functions,
\item all previously defined function macros and
\item all previously declared synthesis functions.
\end{enumerate}

\subsection{Initiating synthesis and synthesizer output \label{sub:Spec:CheckSynth}
\hfill{} $\protect\checksynthcmd$}

\begin{alignat*}{1}
 & \begin{array}{rcl}
\checksynthcmd & ::= & \paren{\checksynthkwd}\end{array}
\end{alignat*}

Synthesis is initiated with $\checksynthcmd$. Exactly those synthesis
functions declared before the occurrence of this command need to be
synthesized. Exactly those constraints occurring before this command
should be satisfied. On successful completion of synthesis, the synthesizer
prints, for each previously declared synthesis function, a well-typed
$\fundefcmd$ drawn from the appropriate syntax, so that all synthesized
functions together satisfy the specification. Otherwise, the synthesizer
prints ${\tt (fail)}$. We give an example of the output produced
by a valid synthesizer on successfully synthesizing the specification
of figure \ref{fig:Example:Max2} in figure \ref{fig:Spec:CheckSynth:Output}.

\begin{figure}
\begin{lstlisting}[basicstyle={\ttfamily}]
(define-fun max2 ((x Int) (y Int)) Int
    (ite (<= x y) y x))

(define-fun min2 ((x Int) (y Int)) Int
    (ite (<= x y) x y))
\end{lstlisting}

\caption{An example of valid synthesizer output to the specification of figure
\ref{fig:Example:Max2}. \label{fig:Spec:CheckSynth:Output}}
\end{figure}

\subsection{Solver-specific options \label{sub:Spec:SetOpts} \hfill{} $\protect\setoptscmd$}

Synthesizer flags and parameters may be controlled with $\setoptscmd$
-- examples include specifying the search strategy, or search parameters
such as expression size. The syntax is as follows: 
\begin{alignat*}{1}
 & \begin{array}{rcl}
\setoptscmd & ::= & \paren{\setoptskwd\mbox{ }\paren{\kplus{\paren{\symbol\mbox{ }\quotedliteral}}}}\end{array}
\end{alignat*}

The behavior of a synthesizer on encountering a $\setoptscmd$ is
implementation defined. It is recommended however, that synthesizers
ignore unrecognized options, and choose reasonable defaults when the
options are left unspecified.

\bibliographystyle{plain}
\bibliography{references}

\begin{thebibliography}{1}

\bibitem{FMCAD13}
Rajeev Alur, Rastislav Bod\'{\i}k, Garvit Juniwal, Milo M.~K. Martin, Mukund
  Raghothaman, Sanjit~A. Seshia, Rishabh Singh, Armando Solar-Lezama, Emina
  Torlak, and Abhishek Udupa.
\newblock Syntax-guided synthesis.
\newblock In {\em FMCAD}, pages 1--17, 2013.

\end{thebibliography}

\end{document}